\documentclass[aps,twocolumn,showpacs,superscriptaddress,showkeys,prd]{revtex4-1} 
\usepackage{dcolumn}   
\usepackage{bm}      
\usepackage{xcolor}
\usepackage[utf8]{inputenc}
\usepackage{amsmath, amssymb,slashed,float}
\usepackage{graphicx}
\usepackage[english]{babel}
\usepackage{hyperref}
\usepackage{tikz}

\newcommand{\beq}{\begin{equation}}
\newcommand{\eeq}{\end{equation}}                     
\newcommand{\beqa}{\begin{eqnarray}}
\newcommand{\eeqa}{\end{eqnarray}}

\graphicspath{{./figures/}}

\usetikzlibrary{decorations.pathmorphing}
\usetikzlibrary{decorations.markings}
\usetikzlibrary{mindmap}
\usepgflibrary[shapes.arrows]
 \definecolor{darkgreen}{RGB}{34,139,34}

\newcommand{\AddrAHEP}{AHEP Group, Institut de F\'{i}sica Corpuscular --
  C.S.I.C./Universitat de Val\`{e}ncia, Parc Cientific de Paterna.\\
  C/Catedratico Jos\'e Beltr\'an, 2 E-46980 Paterna (Val\`{e}ncia) - SPAIN}

\begin{document}

\title{Bounds on Neutrino-Scalar Yukawa Coupling}

\author{P. S. Pasquini}
\email{pasquini@ifi.unicamp.br}
\affiliation{~Instituto de F\'isica Gleb Wataghin - UNICAMP, {13083-859}, Campinas SP, Brazil}
\affiliation{~\AddrAHEP}
\author{O. L. G. Peres}
\email{orlando@ifi.unicamp.br}
\affiliation{~Instituto de F\'isica Gleb Wataghin - UNICAMP, {13083-859}, Campinas SP, Brazil}
\affiliation{~Abdus Salam International Center for Theoretical Physics, Strada Costiera 11, 34014 Trieste, Italy.}

\begin{abstract}
General neutrino-scalar couplings appear in many extensions of Standard Model. We can probe these  neutrino-scalar couplings by leptonic decay of mesons and from heavy neutrino search.  Our analysis improves the present limits to $|g_e|^2<1.9\times 10^{-6}$ and $|g_\mu|^2<1.9\times 10^{-7}$ at 90\% C.L. for massless scalars. For massive scalars we found for the first time the constraints for $g^2_{\alpha}$ couplings to be $10^{-6}-10^{-1}$ respectively for scalar masses between up 100 MeV and we have no limits for masses above 300 MeV.
\end{abstract}

\maketitle

\section{\label{sec:intro}Introduction}
Although in Standard Model there are no couplings between neutrinos and scalar fields due to the non-existence of right-handed neutrino fields, many of its extension contains an enlarged scalar sector that may result in non-universal Yukawa interactions that couples neutrinos and those new scalars.

Non-universal neutrino-scalar couplings can have in\-te\-res\-ting consequences such as:  (i) existence of  new decay channels for particle decays, specially meson decays and lepton decays,~\cite{Barger:1981vd,Lessa:2007up}; (ii) induced neutrino decay~\cite{Zatsepin:1978iy,Gomes:2014yua,Berryman:2014yoa,Picoreti:2015ika,Abrahao:2015rba}; (iii) the presence of new channels for energy loss of supernova caused by enhanced emission of neutrinos and scalars $\chi$~\cite{Farzan:2002wx}, (iv) new channels for neutrinoless double beta decay with the emission of massless $\chi$ in the final state~\cite{Albert:2014fya} and (v) change in flavor ratios of high energy neutrinos from astrophysical sources~\cite{Blum:2014ewa,Dorame:2013lka}.

In general we can parametrize the neutrino-scalar Lagragian to be,
\beq
-{\cal L}=\frac{1}{2}g_{A B}\overline{\nu}_A\nu_B\chi_1+\frac{i}{2}h_{A B}\overline{\nu}_A\gamma_5\nu_B\chi_2,
\label{lag}
\eeq 
where $\chi_1$ ($\chi_2$) is the hypothetical scalar (pseudo-scalar) $\nu_A$ the neutrinos, which may or may not have a right-handed part, and the coupling constant $g_{A B}$ ($h_{A B}$) where $A,B$ runs over two possible basis: (i) $A,B=$ Greek index: neutrino flavour eigenstates $e,\mu$ and $\tau$ and (ii) $A,B=$ Roman index: neutrino mass eigenstates $1,2$ and $3$, that are related by,
\beq
g_{ij}=U^{*}_{i\alpha }U_{\beta j}g_{\alpha \beta}.
\eeq
And equivalently for $h$. 
\def\SM{$\mathrm{SU(3)_c \otimes SU(2)_L \otimes U(1)_Y}$ }
\def\SMextra{$\mathrm{SU(3)_c \otimes SU(2)_L \otimes U(1)_Y \otimes U(1)_H}$ }
\def\SMtres{$\mathrm{SU(3)_c \otimes SU(3)_L \otimes U(1)_N}$ }

Most modern experiments that can probe neutrino-scalar interaction can't distinguish between neutrino final state, so it is convenient to define an effective coupling constant squared:
\beq
|g_l|^2~\equiv ~\sum_{\alpha} \left(|g_{l\alpha}|^2+|h_{l\alpha}|^2\right)
\label{gmassa}
\eeq
with $\alpha,l=e,\mu,\tau$, where U is is the mixing matrix of three lightest neutrinos.

Previous constraints on these couplings are $|g_e|^2<4.4\times 10^{-5}$, $|g_\mu|^2<3.6\times 10^{-4}$ and $|g_\tau|^2<2.2\times 10^{-1}$ at 90\% C.L. and were obtained from meson as well as from lepton decays analysis~\cite{Lessa:2007up}. 

For the absence of detection of neutrinoless double beta decay it was found that $|g_e|^2<\left(0.8-1.6\right)\times 10^{-5}$  at 90 \% C.L.~\cite{Albert:2014fya},  where the uncertainties came from the computation of nuclear matrix elements of the neutrinoless double beta decay. 
All these limits were made in the limit of massless scalar field $\chi$.  

The effective Lagrangian for neutrino-scalar couplings shown in Eq.~(\ref{lag}) can be embedded in different extensions of Standard Model. The general trend is to have the inclusion of new scalar particles in different representations with non-universal couplings between the different families and also the addition of new sterile neutrino states. 
For instance, Ref.~\cite{Dorame:2013lka} presents a model with a \SMextra symmetry that included as new fields one extra singlet scalar boson and three right-handed neutrinos.  
Another example is the model with a \SMtres symmetry that due to anomaly cancellation requirement has already non-universal couplings~\cite{Cogollo:2008zc}. Recently, a lot of theoretical models with scalars that have vacuum expectation values (vev) that are significantly smaller then the vev of Standard Model Higgs, $v_{\rm SM}=246$~GeV were proposed. Examples of these models involve neutrophilic scalars as in Ref.~\cite{Machado:2015sha} with vev $\sim $~eV and 
models with gauged B-L symmetry~\cite{Machado:2010ui,Machado:2013oza} that also have vev much smaller then the SM vev resulting in small scalar masses, ranging from eV to TeV values.  This open an interesting point to study the consequences of non-universal neutrino-scalar couplings for massive scalar fields $\chi$ that did not have been studied so far.   Then our goal is to revisit the bounds on neutrino-scalar couplings for massless scalars and to compute the bounds  on the neutrino with  light {\it massive} scalars. 

Notice that $\chi_{1,2}$ are not the SM Higgs field, for an analysis of the consequences of the Higgs coupling to neutrinos see~\citep{Pilaftsis:1991ug}. Also, the interaction between neutrino scalar may or may not be reminiscent of a neutrino mass generating mechanism, this have to be analysed case by case.

This paper is organized as follows. In Section~(\ref{sec:mesons}) we discuss the computation of meson decay when we have non-universal neutrino-scalar couplings, then in Section~(\ref{subsec:data}) we discuss the available data for meson decay rate and spectrum. In Section~(\ref{analysis}) we made the analysis and extract the constraints on the neutrino-scalar couplings. In Section~(\ref{compare}) we translated the 
bounds on the mass basis and we conclude in Section~(\ref{conclusion}) summarizing our main results.

\section{\label{sec:mesons} Bounds Using Meson Decay}
The leptonic decay rate of a meson P, $P\rightarrow l+\overline{\nu}_l$ at three level is given by,
\beqa
\Gamma^{0}\left(P\rightarrow l\nu_l\right)&=&\dfrac{G_{\rm{F}}^2 f_{\rm{P}}^2 |V_{\rm{qq'}}|^2 m_P^3}{8\pi}\left(x_l+\alpha-(x_l-\alpha)^2\right) \nonumber \\
      & & \times \lambda^{1/2}(1,x_l,\alpha)
\label{rate2}
\eeqa
where $x_l=\left(\dfrac{m_{\nu_l}}{m_P}\right)^2$ and $\alpha=\left(\dfrac{m_l}{m_P}\right)^2$ where $m_{\nu_l}$ is the neutrino mass and $m_l$ is the lepton mass, $G_{\rm{F}}$ is the Fermi constant, $f_{\rm{P}}$ the meson decay constant, $V_{\rm{qq'}}$ is the corresponding CKM-matrix element of the decay and $\lambda(x,y,z)$ is the well known kinematic triangular function~\cite{Agashe:2014kda}.\\

Precise branching ratio measurements of mesons open a window to probe the scalar interaction due to the fact that it is chiral suppressed, e.g.
for massless neutrinos, $m_{\nu_l}\to 0$, the total rate is proportional to $\alpha \propto m_l$ and is very small when $m_P>>m_l$.\\

The precision of $\pi$ and $K$ meson decay rate requires also the inclusion of electromagnetic and weak radiative corrections that can be parametrized as follows,
\beq
\Gamma\left(P\rightarrow l\nu_l\right)=\Gamma^{0}\left(P\rightarrow l\nu_l\right)S_{\rm EW}\left(1+\alpha_{\rm el}G_{\rm rad}\right)
\label{rate1}
\eeq
The electromagnetic radiative corrections $G_{\rm rad}$ came from short and long range corrections and were computed for pions in  Ref.~\cite{Marciano:1993sh} at one loop level and at 2 loop level for pions and kaons in Ref.~\cite{Cirigliano:2007xi}. A nice review of these computations can be found in Ref.~\cite{Cirigliano:2011ny}. The electroweak radiative corrections $S_{\rm EW}$ are given in Ref.~\cite{Marciano:1993sh}.

\begin{figure}[H]
			\includegraphics[scale=0.9]{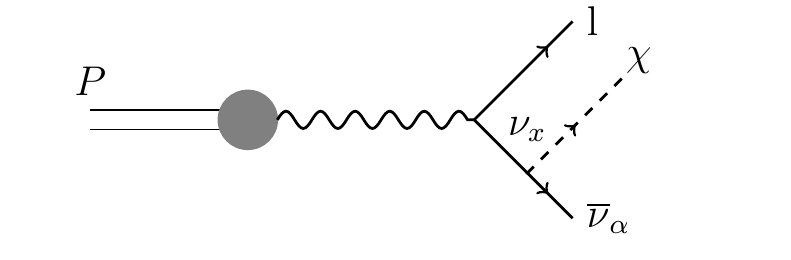}
			\caption{\label{fig0}Tree Level Feynman diagram of the three-body-decay.}
\end{figure}
When we add the interaction from non-universal neutrino-scalar couplings shown in Eq.~(\ref{lag}) the meson have a three body decay
\beq
P\rightarrow l+\overline{\nu}_{\alpha}+\chi
\eeq
where the Feynman diagram is given by Fig.~(\ref{fig0}). The rate was computed in Ref.~\cite{Barger:1981vd}. The differential rate for this process is 
\beq\label{eq:difdecayrate}
d\Gamma\left(P\rightarrow l\nu_\alpha \chi\right)=\Gamma\left(P\rightarrow l\nu_x\right)dR
\eeq
where the rate $\Gamma\left(P\rightarrow l\nu_x\right)$ is similar to the two-body rate of Eq.~(\ref{rate2}) replacing  the mass of real neutrino of the two body decay in final state $\nu_l$ by a virtual neutrino $\nu_x$ of invariant  mass $m_{\nu_x}^2$ as shown in Fig.~(\ref{fig0}). The rate  $\Gamma\left(P\rightarrow l\nu_x\right)$ is given by
\beqa
\Gamma\left(P\rightarrow l\nu_x\right)&=&\frac{G_{\rm{f}}^2 f_{\rm{p}}^2 |V_{\rm{qq'}}|^2 m_P^3}{8\pi}\left(x+\alpha-(x-\alpha)^2\right) \nonumber \\
   & & \times \lambda^{1/2}(1,x,\alpha)
\eeqa
where we made the replacement $x_l\to x=\left(\dfrac{m_{\nu_x}}{m_P}\right)^2$ in Eq.~(\ref{rate2}). To obtain the full rate of the three body decay we should integrate over all  possible invariant mass $m_{\nu_x}$ of the virtual neutrino whose phase space is not a delta fixed by energy conservation but it is an additional independent variable.  The factor dR in Eq.~(\ref{eq:difdecayrate}) is as follows 
\beq\label{eq:phasespace}
dR=\frac{(x^2+\beta^2+6x\beta-\gamma x-\gamma\beta)\lambda^{1/2}(x,\beta,\gamma)}{(x-\beta)^2x^2} \dfrac{|g_{l\alpha}|^2}{32\pi^2}dx
\eeq
where  $\gamma=\left(\dfrac{m_\chi}{m_P}\right)^2$ and the $m_\chi$ is mass of scalar $\chi$,
 and $g_{l\alpha}$ is the coupling of the vertex neutrino-neutrino-scalar $\nu_l-\nu_{\alpha}-\chi$ from Eq.~(\ref{lag}),
The integration limits are $\gamma\leq x\leq (1-\sqrt{\alpha})^2$. Notice that for $\beta,\gamma\rightarrow0$ this integration is infrared (IR) divergent. Previous calculations did not presented a formal treatment of this divergence, Ref.~\cite{Barger:1981vd} assumed $m_ \chi,m_\nu\sim1~\rm{eV}$ and Ref.~\cite{Gelmini:1982rr} took $m_\nu\sim0.1~\rm{MeV}$. We present a finite calculation in the Appendix~\ref{apend:A}.
.

\section{\label{subsec:data} Leptonic Decay Data}
The experimental data used in this work comes from various meson decays measurements, shown in Table~(\ref{tab:data}). The analysis is subdivided into two groups of data:
\begin{itemize}
\item[(a)] The first group comes from rates of  leptonic decay of mesons.
\item[(b)] The second group is obtained from charged lepton spectrum of mesons decay.
\end{itemize} 

\subsection{\label{subsec:rates} Leptonic rates constraints}

Assuming zero neutrino mass (or small enough), and that the experiment can't differentiate between emitted neutrinos, it is possible to write the correction to the decay
rate as,
\beq\label{eq:correction}
\begin{array}{ccc}
\Gamma\left(P\rightarrow\rm{Leptonic}\right)&=&\Gamma_{\rm{SM}}+|g_l|^2\Gamma'
\end{array}
\eeq
where $\Gamma_{\rm{SM}}$ is the expected SM contribution to the P-leptonic decay. $|g_l|^2\Gamma'$ represents the contribution of the  scalar and pseudo-scalar couplings of Eq.~(\ref{lag}),  \beq
|g_l|^2=\sum_{\alpha=e,\mu,\tau} |g_{l\alpha}|^2+|g'_{l\alpha}|^2
\label{gl}
\eeq 
and $~\Gamma'$ is a numerical factor obtained integrating Eq.(\ref{eq:difdecayrate}) that depends on the particle masses.
Notice that if $|g_l|^2\rightarrow0$ we recover SM results.

The data from meson leptonic decay for light mesons ($\pi$ and $K$) include all available space of a three body decay due to the fact that it is hard to separate $P\rightarrow l\nu_l$ and $P\rightarrow l\nu_l \gamma$, thus to obtain the correction $\Gamma'$ we can safely include all $x$ integration limit. This makes the ratio $\frac{\Gamma'}{\Gamma_{\rm{SM}}}$ reaches orders of $10^2-10^3$ due to chiral suppression that combined to the smallness of the experimental error allow us to put stringent bounds from such decays.  

\setlength\extrarowheight{5pt}
\begin{table}[h]
   \begin{tabular}{|c|c|}
  \hline
 Reaction & Reference  \\ \hline\hline
$P\rightarrow l\bar{\nu}$ ($\pi,K$)& \cite{Agashe:2014kda}  \\ \hline
$P\rightarrow l\bar{\nu}$ ($D,D_s,B$)& \cite{Ablikim:2013uvu,Li:2012tr,Lees:2010qj,White:2012zza,Zupanc:2013byn}  \\ \hline
$\pi^+\rightarrow e^+\nu_H$ & \cite{Britton:1992xv}  \\ \hline
$K^+\rightarrow \mu^+\nu_H$ & \cite{Artamonov:2014urb}  \\ \hline
Br($\pi^{+}\rightarrow e^{+}\nu_e\nu\overline{\nu})<5\times10^{-6}$) &  \cite{Agashe:2014kda}\\\hline
Br($K^{+}\rightarrow \mu^{+}\nu_e\nu\overline{\nu})<6\times10^{-6}$)&  \cite{Agashe:2014kda}\\\hline
 \end{tabular}
\caption{Reactions used in this work.}
\label{tab:data}
\end{table}

For heavy mesons  such as $D$, $D_s$ and $B$ the leptonic decay rate of heavy mesons is suffering from large background of hadronic decays and
the measurement of meson decay is triggered by the detection of the charged lepton in the final state and a missing four momentum. In the SM the missing energy comes from the neutrino of the two body decay that we are assuming to be very small which is equivalent to $M^2_{\rm{Miss}}=m^2_{\nu_l} \sim 0$. Nevertheless experiments can only select leptonic events with $M^2_{\rm{Miss}}\lesssim 0.1\rm{GeV}^2$ due to detector limitations. This opens a window to probe scalar masses different from zero, we can related the missing energy with the $x_l$ variable $M^2_{\rm{Miss}}=x_lm_P^2$ with the SM two-body decay. When we include the three body decay the relevant variable is $x$ and we can related $M^2_{\rm{Miss}}=x m_P^2$. Thus the experimental cut  
$\left(M^2_{\rm{Miss}}\right)^{\rm max}\lesssim 0.1~\rm{GeV}^2$ translates to an upper limit in the range of variable x, 
$\gamma\leq x\leq x_{\rm max}$. where $x_{\rm max}=\left(M^2_{\rm{Miss}}\right)^{\rm max}/m_P^2$.

\begin{table}[H]
\centering
   \begin{tabular}{|c|c|c|}
  \hline
   Element & Most Precise & Not from Leptonic Decay\\ \hline\hline
$|V_{\rm{ud}}|$& ------  &  $0.97425(22)$ \\ \hline
$|V_{\rm{us}}|$& $0.2253(10)$& $0.2253(14)$  \\ \hline
$|V_{\rm{cd}}|$& $0.225(8)$& $0.220(12)$  \\ \hline
$|V_{\rm{cs}}|$& $0.986(16)$&$0.953(25)$ \\ \hline
$|V_{\rm{ub}}|$& $4.22(42)\times10^{-3}$& $4.13(49)\times10^{-3}$\\ \hline
 \end{tabular}
\caption{Most precise $V_{\rm{CKM}}$ matrix elements compared to those used here. All of the values comes from the PDG~\cite{Agashe:2014kda}.}
\label{CKM-elements}
\end{table}

Two points should be considered when we compare the data from  meson leptonic decay rate and the theoretical predictions: the value of meson constant $f_P$ and of the CKM elements $|V_{\rm{qq'}}|^2$. Both quantities usually were obtained from the two-body leptonic decay (See for example, Ref.~\cite{Agashe:2014kda}) and then if we want to test the two-body for new physics we cannot use these values. In previous references~\cite{Barger:1981vd,Lessa:2007up} it was decided to use the $f_P$ and $|V_{\rm{qq'}}|^2$ due to the fact that we cannot extract the constraints to neutrino-scalar couplings  without these assumption. In this work we decide to use the information for $f_P$ and $|V_{\rm{qq'}}|^2$ from other places in the following way:
\begin{itemize}
\item Some precise measurements of the CKM matrix elements ($D, D_s$ and $B$) comes exactly from the fit of meson leptonic decay rate measurements~\cite{Agashe:2014kda}, thus those results could be contaminated by exactly the decays we want to find. The solution is to use other measurements of CKM mixing matrix, such as the meson beta decay (e.g. $D\rightarrow\pi l\nu$) and pay the price of less precise values. Table~\ref{CKM-elements} compares the most precise values of CKM mixing matrix and the ones use here that don't come from leptonic decays.
\begin{table}[H]
\centering
   \begin{tabular}{|c|c|}
  \hline
   & $f_P [\rm{MeV}]$ \\ \hline\hline
$\pi$& $130.2(1.4)$     \\ \hline
$K$& $156.3(0.9)$     \\ \hline
$D$& $209(3.3)$     \\ \hline
$D_s$& $250(7)$     \\ \hline
$B$& $186(4)$     \\ \hline
 \end{tabular}
\caption{Form Factors $f_p$ from Lattice QCD~\cite{Aoki:2013ldr}}
\label{formfactors}
\end{table}
\item Meson decay constant, $f_P$, is also measured from leptonic decay for light mesons ($\pi$ and $K$). Recently lattice QCD was able to obtain them with good precision, enabling this kind of analysis for the first time, the numerical values of $f_P$ can be found in Ref.~\cite{Aoki:2013ldr} and we listed in Table~\ref{formfactors}.

\end{itemize}

\subsection{\label{subsec:spectrum} Leptonic spectrum constraints}

The second group is obtained from heavy neutrino search which scans the charged lepton spectrum for peaks of two-body decays. This is the first time that this kind of data is used to restrict Neutrino-Scalar couplings. We obtain the best constraints on these couplings from the spectrum analysis. 

\begin{figure}[H]
 		\includegraphics[height=4cm,width=0.5\textwidth]{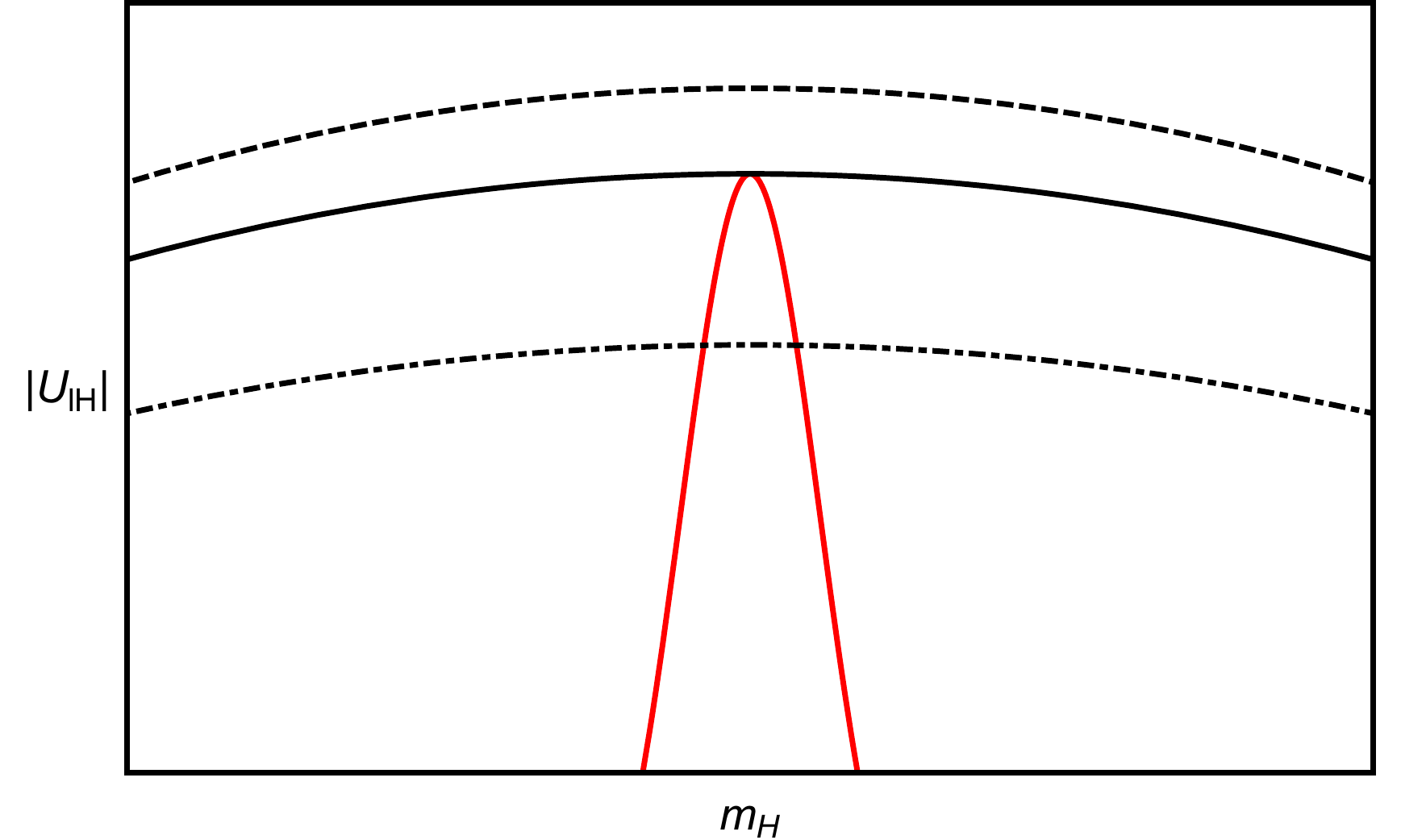}
		\caption{\label{fig0a} This plot shows three hypothetical scenarios, the red line represents the peak search, the dashed line a signal and the dotted-dashed a negative signal, the solid line is the limiting case.}
\end{figure}

Inspection of light meson decay spectrum was used to search for heavy neutrinos, of mass $m_H$, by peak search, in special~\cite{Britton:1992xv,Artamonov:2014urb} found no evidence, putting bounds on the heavy neutrino mass and its mixing matrix to the active neutrinos. The contribution to the charged lepton spectrum can be parametrize on the following form~\cite{Artamonov:2014urb},
\beq\label{eq:heavy}
d\Gamma(P\rightarrow l\nu_H)=\rho\Gamma_0 |U_{eH}|^2\delta(p_{\rm{peak}}-p_l)dp_l,
\eeq
With $\Gamma_0$ from equation Eq.~(\ref{rate2}) setting $m_\nu=0$,  $|U_{eH}|^2$ is the mixing of the heavy neutrino presented in the decay $P\to e +\nu_H$,  $p_l$ is charged lepton momentum. The Heavy neutrino mass information is contained only in $p_{\rm peak}$, the charged lepton momentum  expected for two-body decay of meson P,
\beq
p_{\rm{peak}}=\frac{\lambda^{1/2}(m_P^2,m_l^2,m_H^2)}{2m_P},
\eeq 
and in the variable $\rho$ that is given by
\beq
\rho=\frac{\sqrt{1+(\alpha-\beta)^2-2(\alpha+\beta)}(\alpha+\beta-(\alpha-\beta)^2)}{\alpha(1-\alpha)^2}
\eeq
which is the correction for the two-body meson decay for a heavy neutrino H with mass $m_H$ and
 $\beta=\left(\frac{m_H}{m_P}\right)^2$ compared with a massless neutrino limit of Eq.~(\ref{rate2}).

We can use this information to constrain the neutrino-scalar couplings in the following way. The spectrum with an heavy neutrino should have a peak at mass of heavy neutrino $m_H$ proportional to  the mixing of heavy neutrino  $|U_{eH}|^2$ with the electron neutrino (See Fig.~(\ref{fig0a})). The three body decay from  Feynman diagram shown in Fig.~(\ref{fig0}) have a continuous spectrum show in Fig.~(\ref{fig0a}) by the dashed, solid and dotted-dashed curves  respectively that have numerical values above, equal and below the maximum values of spectrum of heavy neutrino in two body decays.  Saying in other words, we can put a bound by comparing the number of events in peak search area (below the two-body heavy neutrino search) and the three body search.  Effectively  the rate of hevay meson decay given in   Eq.~(\ref{eq:heavy}) to be equal to the three body decay rate given in   Eq.~(\ref{eq:difdecayrate}):
\beq\label{eq:boundU}
|U_{l H}|^2=\left.\dfrac{\Gamma(P\rightarrow l\nu_x)}{\rho(\alpha,x,\beta)\Gamma_0}\dfrac{dR}{dp_l}\right|_{\beta\rightarrow x}
\eeq
and using the constraints from heavy neutrino that constrain the variables $|U_{l H}|^2\times m_h$~\cite{Artamonov:2014urb} we can get constraints on neutrino-scalar couplings.

\section{Analysis and Results}
\label{analysis}

We are going to get the bounds from neutrino-scalar couplings for different values of scalar particle. First we are going to do the case studied so far in the literature~\cite{Barger:1981vd,Lessa:2007up}, when the scalar have zero mass, $m_{\chi}\to0$ in Section.~(\ref{subsec:1}) and in the Section~(\ref{subsec:2}) for the case of $m_{\chi}\neq 0$.

\subsection{Case I: $m_\chi=0$}
\label{subsec:1}
To obtain bounds on the Yukawa coupling constants we used a $\chi^2$ method, defining it as,
\beq\label{eq:chisquared}
\chi^2=\sum_i \frac{\left(\Gamma_{\rm{Teo}}^{(i)}-\Gamma_{\rm{Exp}}^{(i)}\right)^2}{\sigma_i^2}
\eeq
where $i$ run over the experimental data points. To obtain those bounds we marginalized all three CKM elements ($V_{\rm{us}},V_{\rm{cd}},V_{\rm{cs}}$) by varying four parameters: $|V_{\rm{CKM}}|$'s (and a coupling constant $|g_l|^2$ to find the $\chi^2_{\rm{min}}$. Our results on $m_\chi=0$ can be compared with previous bonds found on literature on Table~(\ref{tab:bounds1}).  It is possible to see that in this scenario the tau coupling constant is poorly constrained due to the fact that the errors from the mesons decays and CKM matrix are rather large.

\begin{table}[H]
   \begin{tabular}{|c|c|c|c|}
  \hline
 Constants &Ref.~\cite{Lessa:2007up} & Ref.~\cite{Albert:2014fya} & Our Results   \\ \hline\hline
$|g_e|^2$  & $<4.4\times10^{-5}$   & $<\left(0.8-1.6\right)\times 10^{-5}$ & $<4.4~{\color{red} (4.4)}\times10^{-5}$ \\ \hline
$|g_\mu|^2$  &$<3.6\times10^{-4}$ & &  $<4.5~{\color{red} (3.6)}\times10^{-6}$  \\ \hline
$|g_\tau|^2$  &$<2.2\times10^{-1}$ & &  $<40~{\color{red} (8)}$  \\ \hline
 \end{tabular}
\caption{Comparison between previous bounds~\cite{Lessa:2007up,Albert:2014fya} with our results with $m_\chi=0$, using  the rates of the meson decay  at $90\%$  C.L. In Black the bounds marginalizing $V_{\rm{CKM}}$ in Red, taking the central value of uncorrelated measurements.}
\label{tab:bounds1}
\end{table}
 In contrast the case of assuming fixed the central value of $|V_{\rm{CKM}}|$ using the second column of Table~\ref{CKM-elements} we can bound it as $|g_\tau|^2< 8$, which is still bigger than the previous bounds obtained from tau decay~\cite{Lessa:2007up}. The other coupling constants with fixed CKM are discribed in red in Table~\ref{tab:bounds1}. Notice that our analysis is more complete than those from the literature that have not taken into account the possible correlation between the measurements from $|V_{\rm{CKM}}|$ and the bounds on $|g_l|^2$.
\begin{table}[H]
   \begin{tabular}{|c|c|c|c|}
  \hline
 Constants   & Ref.~\cite{Lessa:2007up} & Ref.~\cite{Albert:2014fya} &Our Results\\ \hline\hline
$|g_e|^2$& $<4.4\times10^{-5}$   & $<\left(0.8-1.6\right)\times 10^{-5}$ & $<1.9\times10^{-6}$   \\ \hline
$|g_\mu|^2$  &$<3.6\times10^{-4}$ & &  $<1.9\times10^{-7}$ \\ \hline
 \end{tabular}
\caption{Comparison between previous bounds~\cite{Lessa:2007up,Albert:2014fya} with our results at $m_\chi=0$ with Meson Decay rate  and lepton spectrum from heavy neutrino search at $90\%$ C.L..}
\label{tab:bounds2}
\end{table}
From the Equation~(\ref{eq:boundU})  we can get the constraints from  Refs.~\cite{Britton:1992xv,Artamonov:2014urb} and translate to bounds on neutrino-scalar couplings.
\begin{figure}[H]
   \includegraphics[height=5.cm,width=0.4\textwidth]{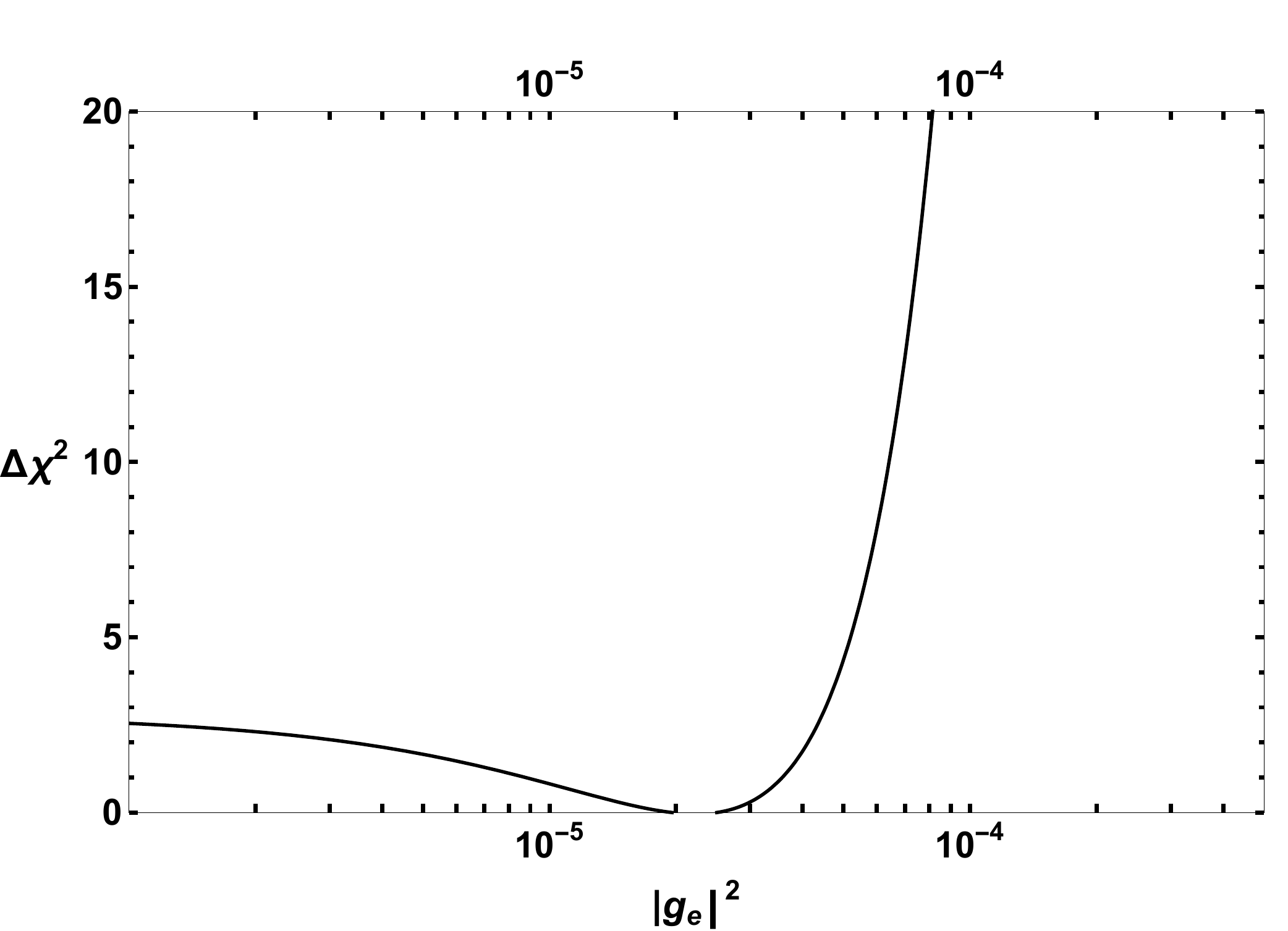} 
   \caption{Marginalized $\Delta\chi^2$ as a function of the value of $|g_e|^2$.}
  \label{marginalizes}
\end{figure}
The data from heavy neutrino search can be used to put bounds on the coupling constants too, all results are summarized on Table~\ref{tab:bounds2}. One can see that heavy neutrino search are one to three orders of magnitude more stringent than those from branching ratios, since it takes into account the decay spectrum.
\begin{table}[hbt]
 \begin{minipage}[c]{0.28\textwidth}
 \begin{tabular}{|c|c|c|}
  \hline
 Parameter & $\ln[B]$  \\ \hline\hline
$|g_e|^2$& $5.7$   \\ \hline
$|g_\mu|^2$ & $7.2$  \\ \hline
$|g_\tau|^2$ & $1.5$\\ \hline
\end{tabular}
	  \end{minipage}
	  \begin{minipage}[c]{0.19\textwidth}
\caption{\label{tab:bayes}Bayes Factor of SM+Scalar over SM.}
		\end{minipage}
\end{table}
The meson decay analysis shows that the SM+Scalar has a minimum $\Delta\chi^2$ for $|g_l|^2\neq0$ which can be seen for $l=e$ case in Figure~\ref{marginalizes}. Thus, to evaluate whether or not the SM+Scalar with $|g_l|^2\neq0$ is a better model than assuming only the SM we compared both situations by using Bayesian inference taking the prior as a normalized Gaussian distribution around each experimental point as evidence $p$,
\beq
p(g|data,M_{\rm{SM}+\chi})=Ne^{-\frac{1}{2}\chi^2(g^2)}
\eeq
then using the Bayes factor, which can be defined as~\cite{Trotta:2008qt},
\beq\label{eq:bayes}
B=\frac{p(d|M_{\rm{SM}})}{p(d|M_{\rm{SM}+\chi})}=\left.\frac{p(g^2|d,M_{\rm{SM}+\chi})}{p(g^2|M_{\rm{SM}+\chi})}\right|_{g^2=0},
\eeq
we can compare the models by calculating $B$: (I) if $B<<1$ the model with more parameters is favored over the  model with less parameters, on the other hand, (II) if $B>>1$ the  model with less parameters is favored due  to describe the data. (III) If $B\approx1$ the data points do not contribute significantly to distinguish between both models.
The last equality in Eq.(\ref{eq:bayes}) is true when the model $M_\chi$ has one extra parameter, $g$ compared to $M_{\rm{SM}}$ and reduces to the same results when $g^2=0$, then, $p(g^2|d,M)$ is the probability of the value $g^2$ of the parameter given the data points and assuming $M$ true and $p(g^2|M)$ is the probability distribution of $g^2$ in the model, in this case $p(g^2|M_{\chi})=(4\pi)^{-1}$ for $\frac{g^2}{4\pi}<1$ and zero otherwise.  We have found  the $B$ values for  $|g_i|^2\neq 0 $ and we shown in Table~\ref{tab:bayes}. For the three couplings constants the preference of SM+scalar model over the SM is less then $2\sigma$ away and from this we conclude there is no stronger preference for $|g_i|^2\neq 0 $.
\begin{figure}[H]
   \includegraphics[height=6cm,width=0.5\textwidth]{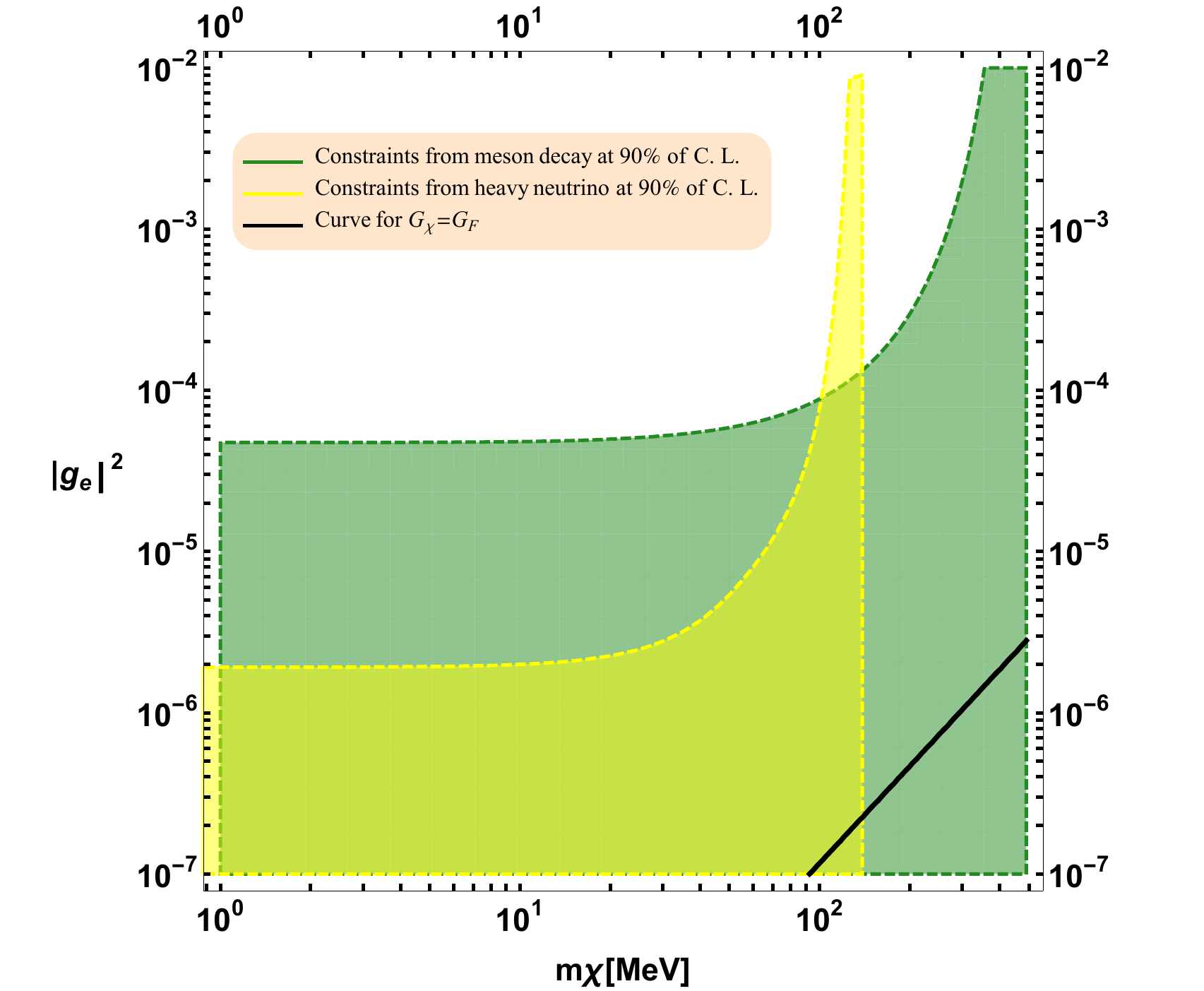} 
   \caption{Bounds on $|g_e|^2$ versus $m_\chi$. The Blue curve comes from heavy neutrino and the red curve comes from meson decay at $90\% \rm{C.L.}$.}
  \label{fig1:Bounds}
\end{figure}

\subsection{Case II: $m_\chi\ne0$}
\label{subsec:2}
This case was never studied and correspond to the general case when $m_\chi\ne0$. 
We proceeded similarly with Section~(\ref{subsec:1}), but using the central value from Table~\ref{CKM-elements} for the CKM mixing elements and thus using Eq.~(\ref{eq:phasespace}).  Now we have two independent variables from the neutrino-scalar lagrangian, the $m_{\chi}$ and the couplings $|g_l|^2$ as defined in Eq.~(\ref{gl}).
\begin{figure}[h]
\includegraphics[height=6cm,width=0.5\textwidth]{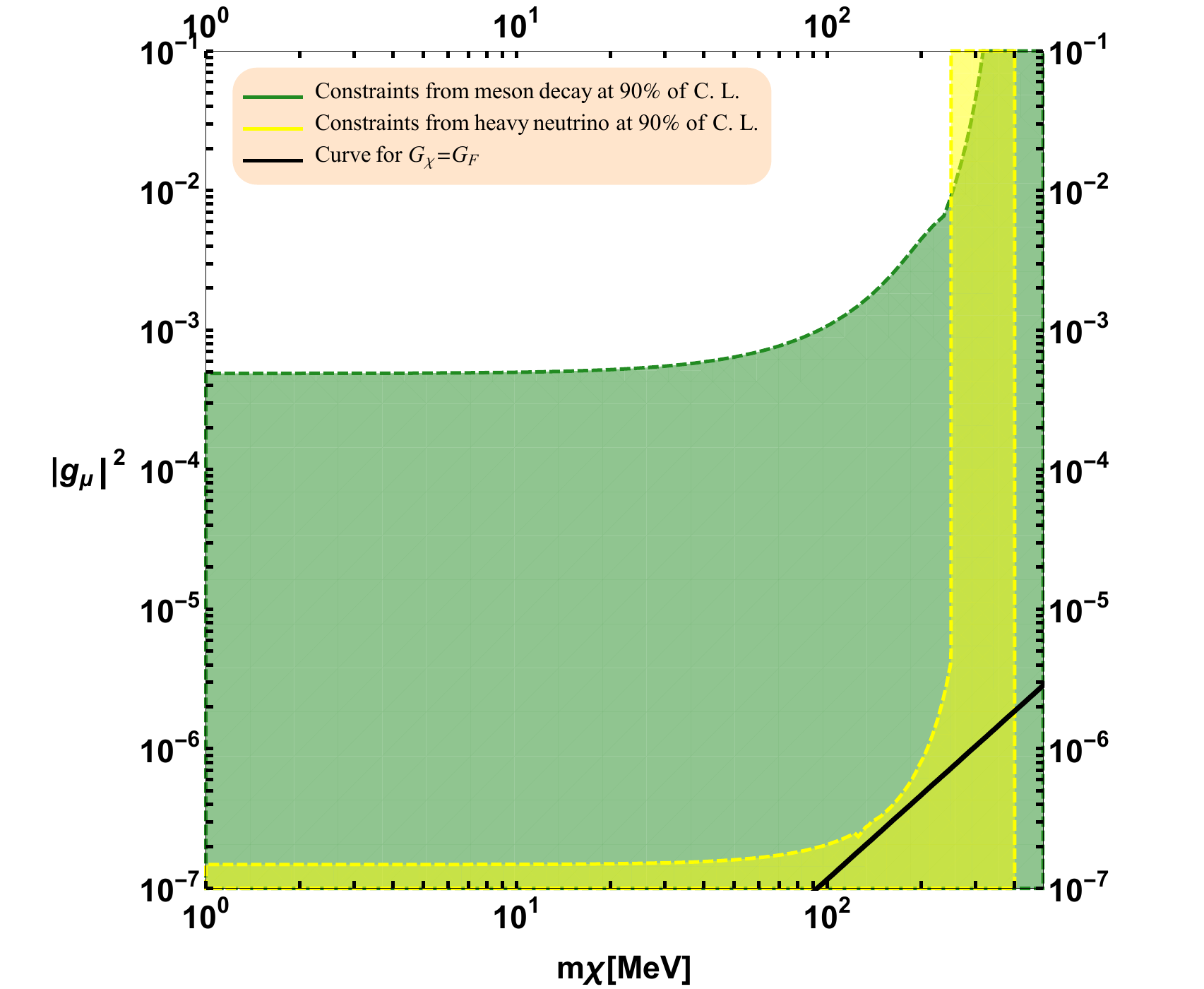} 
\caption{Bounds on $|g_\mu|^2$ versus $m_\chi$. The yellow  curve comes from heavy neutrino search in lepton spectrum and the  green curve comes from rate of meson decay at $90\%$ C.L..}
\label{fig2:Bounds}
\end{figure}
Our results are shown in Figure~(\ref{fig1:Bounds}),(\ref{fig2:Bounds}) and~(\ref{fig3:Bounds}), 
respectively for  $|g_e|^2$, $|g_{\mu}|^2$ and $|g_{\tau}|^2$ for the constraints from the rate of leptonic meson decay and the lepton spectrum respectively in the green and yellow curves at 90 \% C.L.
Notice that in both cases $|g_e|^2$ and $|g_\mu|^2$ the bounds can be assumed to be constant up to masses of order of $\sim200~\rm{MeV}$ and $\sim 100~\rm{MeV}$ respectively.
The constraints for $|g_{\tau}|^2$  are weaker due low statistics of experimental data and also the larger is the lepton mass less effective is the chiral suppression of the two-body meson decay.
\begin{figure}[hbt]
		   \includegraphics[height=5cm,width=0.4\textwidth]{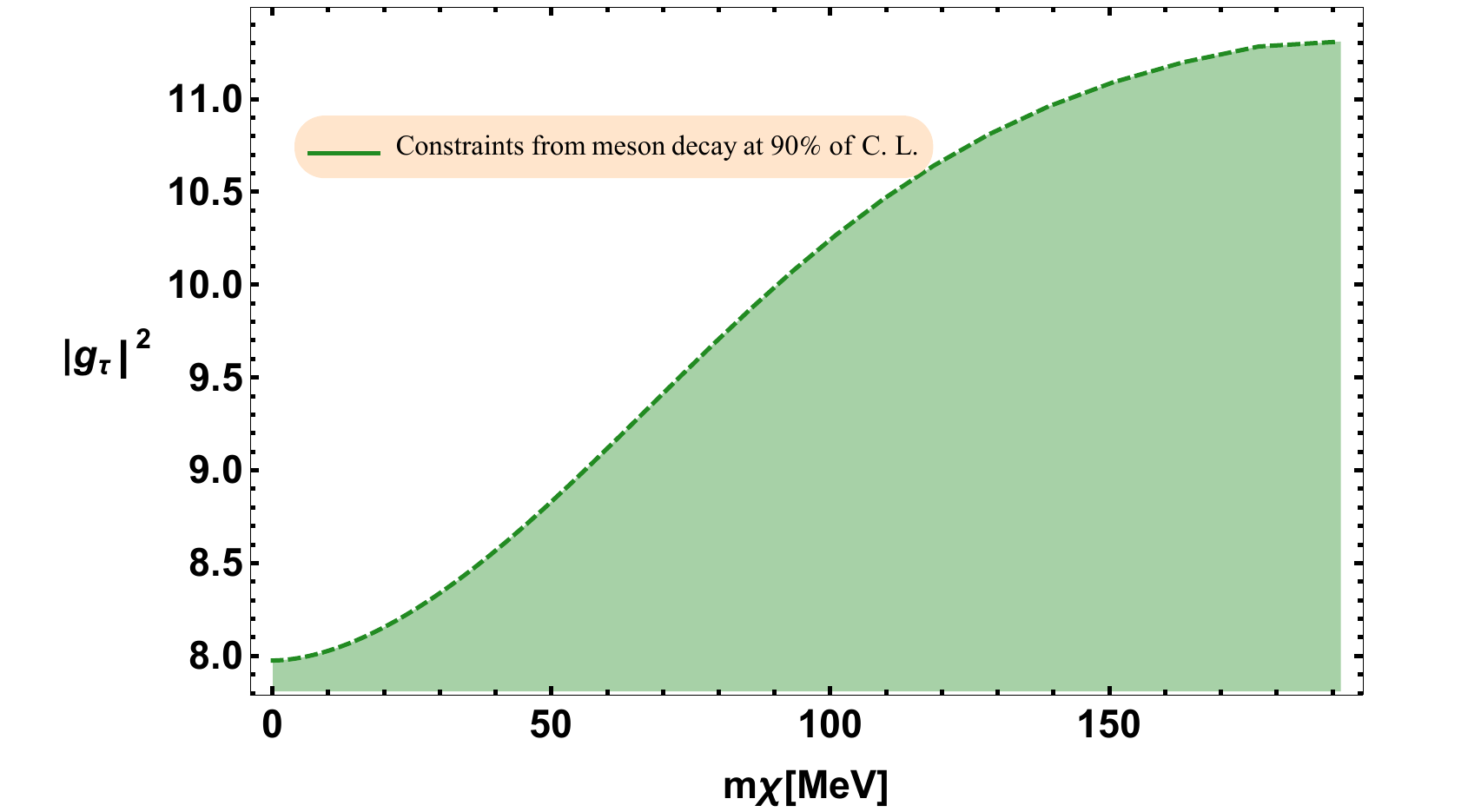} 
   \caption{Bounds on $|g_{\tau}|^2$ versus $m_\chi$,the red curve comes from meson decay at $90\%$ C.L. .}
  \label{fig3:Bounds}
\end{figure}
To have an intuitive idea of size of our constraints for the case of $m_\chi\ne0$ we can compare the strengh of neutrino-scalar interaction represented by $G_\chi \equiv |g_{l}|^2/m_{\chi}^2$, $l=e,\mu,\tau$  with the strenght of the weak interaction $G_F$, they are equal when
\beq
G_\chi=\frac{|g_{l}|^2}{m_\chi^2}\leq G_{\rm{F}}
\eeq
where $l=e,\mu,\tau$ and the value of $ G_{\rm{F}}$ is taken from \cite{Agashe:2014kda}.
We shown in  Figure~\ref{fig1:Bounds} and ~\ref{fig2:Bounds}  the black curve show this equality and any  value of $|g_l|^2$ and $m_{\chi}$ above this curve it is stronger then weak interaction.
\subsection{Higgs Decay}
\label{subsec:3}

One important consequence that may arise from the existence of low mass scalar particles is that it may induce invisible decays of the Higgs field. New LHC that constraint such decay at Br$(h\rightarrow\rm{invisible})<12\%$~\cite{Bernon:2014vta,Belanger:2015kga,Bonilla:2015uwa}.

In general, those invisible decays comes from couplings of the form
\beq
{\cal L}=-\frac{\lambda}{4}|H|^2|\chi|^2
\eeq
that allow three point interactions like $h|\chi|^2$ after symmetry breaking $H\rightarrow h+v$. This implies that $\lambda v\leq m_\chi$ and Br$(h\rightarrow\chi\chi)\leq m_\chi^2/32\pi m_h$ which for current values of the Higgs mass and scalar mass sensitivity imply Br$(h\rightarrow\chi\chi)<0.47\%$, below present experimental accuracy

\section{Comparison with Neutrino decay Bounds}\label{compare}

This non-universal coupling can induce neutrino decays~\cite{Zatsepin:1978iy} that have an interesting phenomenology to made seasonal changes in  solar neutrinos rate~\cite{Picoreti:2015ika} and affects neutrino oscillation behavior  from longbaseline experiments~\cite{Gomes:2014yua} such as MINOS~\cite{Adamson:2011ig} and T2K~\cite{Abe:2014ugx}.  %

From  Eq.~(\ref{lag})  and assuming that the scalar mass is tiny enough, the mass eigenstate neutrinos can decay into one another during propagation via the decay $\nu_i \to \nu_j + \chi$. The neutrino lifetime from such decay was computed by~\cite{Kim:1990km} assuming the third mass eigenstate to be much heavier than the light ones, $m_3>>m_{\rm{light}}$,
\begin{eqnarray}
\dfrac{\tau_3}{m_3}=\dfrac{128\pi}{\left(\sum |g_{3j}|^2+|h_{3j}|^2\right)m_3^2}
\label{eq:majoron}
\end{eqnarray}
The Reference~\cite{Gomes:2014yua}  analyze  MINOS and T2K experiments and compared the expected flux only from oscillated neutrinos and oscillated neutrinos plus decay to obtain a bound on the  neutrino lifetime of $\tau_3/m_3 > 2.8 \times 10^{-12}~{\rm s/eV}$ at $90\%$ C.L. Inserting this result in Eq.~(\ref{eq:majoron}) the obtained bound can be translated to
\begin{eqnarray}
\sum_{j=1,2}|g_{3j}|^2+|h_{3j}|^2<3\times 10^{-2}\left(\dfrac{\rm{eV}}{m_3}\right)^2,
\label{g3}
\end{eqnarray}
where g and h are respectively the neutrino couplings in the mass basis with scalars  and pseudo scalars.  This limit is independent of  mass neutrino hierarchy between the states 2 and 3.

Another analysis taking into account the decay effects on solar neutrino data was performed by~\cite{Picoreti:2015ika} whose neutrino life time obtained was $\tau_2/m_2\geq7.2\times10^{-4}~\rm{s}.\rm{eV}^{-1}$ at 90\% C.L. which would give a bound of,
\beq
|g_{21}|^2+|h_{21}|^2<  1.5\times 10^{-5}\left(\dfrac{\rm{eV}}{m_2}\right)^2.
\label{g2}
\eeq
at 90\% C.L.

Our bounds listed in Table~\ref{tab:bounds1} and Table~\ref{tab:bounds2} are in neutrino-scalar  flavor basis. We can translated these bounds to neutrino mass basis using the relation of Eq.~(\ref{gmassa}). From the analysis of neutrino experiments~\cite{Gonzalez-Garcia:2014bfa} we can have the allowed range for the matrix elements of the mixing matriz U.  First  we assume the bounds for  $|g_{e}|^2$ and/or  $|g_{\mu}|^2$ from Table~(\ref{tab:bounds1}) and (\ref{tab:bounds2}) are valid then
\begin{equation}
|g_{1j}|^2<3\times 10^{-6}\quad |g_{2j}|^2< 4\times 10^{-7}\quad |g_{3j}|^2<5\times 10^{-7}
\label{g1grande}
\end{equation}
where j=1,2,3; otherwise if we assume that only the bounds on $|g_{\tau}|^2$ are valid then the limits are much more weaker: 
 $|g_{1j}|^2<7\times 10^{-3}\quad |g_{2j}|^2< 2\times 10^{-1}\quad |g_{3j}|^2<1\times 10^{-1}$. 
The constraints from neutrino decay in Eq.~(\ref{g2}) and (\ref{g3}) are dependent of the mass of heavier neutrino mass eigenstate.   For the degenerate mass scenario of $m_3\sim m_2 \sim 1$~eV, the constraints from neutrino decay, in Eq.~(\ref{g2}) and (\ref{g3}) are always less restrictive then the results of this work, Eq.~(\ref{g1grande}).

\section{Conclusion}
\label{conclusion}

We compute the bounds for Yukawa interactions between Neutrinos and Hypothetical scalar particles $\chi$ using recent data and decay rates , (I) $m_\chi=0$ and obtaining the neutrino-scalar (pseudo-scalar) couplings in the flavor basis $|g_e|^2<1.9\times10^{-6}$, $|g_\mu|^2<1.9\times10^{-7}$ at $90\%$ C.L., which is an improvement on previous results in literature and {\em for the first time}  (II) $m_\chi\neq0$ showing that those bounds for  $m_\chi=0$ can be safely used up to $100~\rm{MeV}$ scales and no bounds can be put for masses $m_\chi\gtrsim 300~\rm{MeV}$.

For the mass basis the upper bound on neutrino-scalar couplings are $|g_{1j}|^2<3\times 10^{-6}$, $|g_{2j}|^2< 4\times 10^{-7}$ and $|g_{3j}|^2<5\times 10^{-7}$ that much better then the indirect constrain from neutrino decay. 

We can conclude that  we have no evidence for non-universal couplings between neutrino and scalar (pseudo-scalar) and we get the best bounds from the meson decay rate and spectrum data.
\section{Acknowledgments}
 O. L. G. P thanks the hospitality of IPM-Iran and the support of FAPESP funding grant 2012/16389-1. P. S. P. thanks  the support of FAPESP funding grant 2014/05133-1 and 2015/16809-9.

\appendix
\section{IR Treatment}\label{apend:A}
The differential rate given by Eq.~(\ref{eq:difdecayrate}) is Infrared (IR) divergent, as can be seen by expanding it for small $x$ and $m_\chi,\beta\rightarrow0$,
\beq\label{diverg}
d\Gamma(P\rightarrow l\nu_\alpha \chi)\stackrel{x\rightarrow \beta}{\rightarrow}\Gamma_0\frac{|g_l|^2}{32\pi^2}\frac{1}{x}dx
\eeq
 which integrated goes as $\log(x)$ being divergent for the integration limits $0<x<(1-\sqrt{\alpha})^2$.
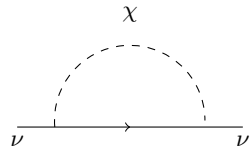
\begin{figure}[H]
  \begin{minipage}[c]{0.25\textwidth}
		\begin{tikzpicture}[scale=1,every node/.style={scale=1}]
			\begin{scope}[decoration={markings,mark=at position 0.5 with {\arrow{>}}}]
\draw[postaction=decorate] (0,0) -- +(3,0) node[anchor=north] {$\nu$} node [at start, anchor=north] {$\nu$};
\end{scope}
\draw[dashed] (.5,0) arc (185:0: 1 and 1);
\node at (1.5,1.5) {$\chi$};
			\end{tikzpicture}
	  \end{minipage}
	  \begin{minipage}[c]{0.2\textwidth}
			\caption{\label{selfenergy} Neutrino Self-Energy.}
		\end{minipage}
\end{figure} 
On the other hand, the correction of the neutrino pro\-pa\-ga\-tor given by the diagram of Figure~\ref{selfenergy}
was calculated by Ref.~\cite{Schiopu:2007zz} and changes the normalization of the neutrino field, $\delta Z$  by,
\beq
\delta Z=-\frac{g^2}{32\pi^2}B_0(p;m_\chi,m_\nu)
\eeq
where $B_0$ is one of the Passarino-Veltman Functions. This induces a change in the completeness relation,
\beq
\sum_s u(p,s)\overline{u}(p,s)=\slashed{p}(1+\delta Z)+O(m_\nu)
\eeq
here, $p$ is the neutrino four-momentum. Thus, one should add to the two-body decay $P\rightarrow l\nu_l$ the renormalization correction,
\beq\label{divergminus}
\Gamma(P\rightarrow l\nu)=\Gamma_0\left(1-\frac{|g_l|^2}{32\pi^2}B_0(p,m_\chi,m_\nu)\right)
\eeq
 Expanding $B_0$ around zero neutrino mass, one gets,
\beq
B_0(p,m_\chi,m_\nu)\stackrel{m_\nu\rightarrow0}{\rightarrow}\log\left(\frac{E^2}{m_\chi^2}\right)=\log\left[\frac{(m_P^2-m_l^2)^2}{4m_{\chi}^2m_P^2}\right]
\eeq
but,
\beq
\log\left[\frac{(m_P^2-m_l^2)^2}{4m_{\chi}^2m_P^2}\right]=\log\left[\frac{(m_P+m_l)^2}{4m_P^2}\right]+\int_\gamma^{(1-\sqrt{\alpha})^2}{\frac{dx}{x}}
\eeq
Thus, summing both contributions the $1/x$ is canceled at small $m_\chi$ due to opposite signs between the corrections in Eq.~(\ref{diverg}) and Eq.~(\ref{divergminus}).

\bibliography{Article_Majoron_2015} 

\end{document}